\documentclass[%
 reprint,
superscriptaddress,
 amsmath,amssymb,
 aps,
]{revtex4-1}
\usepackage{xcolor}
\usepackage{graphicx}
\usepackage{booktabs}
\usepackage{amsmath}
\usepackage{mathrsfs}
\usepackage{graphicx}
\usepackage{caption}
\usepackage{subcaption}
\usepackage{dcolumn}
\usepackage{bm}
\usepackage{hyperref}

\begin{document}

\preprint{APS/123-QED}

\title{Testing black holes in a perfect fluid dark matter environment using quasinormal modes
}%

\author{Qun Tan}
\affiliation{%
 College of Physics,Guizhou University,Guiyang,550025,China
}%

\author{Dong Liu}
\affiliation{Department of Physics, Guizhou Minzu University, Guiyang, 550025, China}

\author{Jie Liang}

\author{Zheng-Wen Long}%
\email{zwlong@gzu.edu.cn (corresponding author)}
\affiliation{%
 College of Physics,Guizhou University,Guiyang,550025,China
}%





\date{\today}

\begin{abstract}
This research explores the quasinormal modes (QNMs) characteristics of charged black holes in a perfect fluid dark matter (PFDM) environment. Based on the Event Horizon Telescope (EHT) observations of the M87* black hole shadow, we implemented necessary constraints on the parameter space($a/M$,$\lambda/M$).We found that for lower values of the magnetic charge parameter, the effective range of the PFDM parameter is approximately between $-0.2$ and $0$, while as the magnetic charge parameter increases, this effective range gradually extends toward more negative values. Then through sixth-order WKB method and time-domain method, we systematically analyzed the quasinormal oscillation spectra under scalar field and electromagnetic field perturbations. The results reveal that: the magnetic charge $a$ and PFDM parameters $\lambda$ modulate the effective potential barrier of black hole spacetime, profoundly influencing the response frequency and energy dissipation characteristics of external perturbations. The consistently negative imaginary part of QNMs across the entire physical parameter domain substantiates the dynamical stability of the investigated system. Moreover, we discovered differences in the parameter variation sensitivity between scalar field and electromagnetic field perturbations, providing a theoretical basis for distinguishing different field disturbances. These results not only unveil the modulation mechanisms of electromagnetic interactions and dark matter distribution on black hole spacetime structures but also offer potential observational evidence for future gravitational wave detection and black hole environment identification.
\begin{description}
\item[Keywords]quasinormal modes,perfect fluid dark matter,black hole shadow,magnetic charge

\end{description}
\end{abstract}

\maketitle

\section{Introduction}\label{1.0}
Black holes, as a theoretical prediction of General Relativity(GR), have now been confirmed through multiple observational pieces of evidence, verifying their existence in the actual universe. The first direct detection of gravitational waves from a binary black hole merger event GW150914 by the LIGO Gravitational Wave Observatory in 2015\cite{LIGOScientific:2016emj, LIGOScientific:2016vlm}, along with the first direct imaging of the supermassive black hole M87* by the EHT\cite{EventHorizonTelescope:2019dse}, provide direct and definitive evidence for black holes. These milestone observations not only validate the fundamental predictions of Einstein's GR but also transform black holes into an ideal natural laboratory for studying strong gravitational field spacetime geometry and physical characteristics, providing observational foundations for exploring frontier topics such as gravitational lensing effects, photon ring structures, gravitational theory verification, and black hole perturbation theory\cite{Virbhadra:1999nm,Gralla:2019xty,Zhao:2024lts,Konoplya:2011qq,Liang:2024geh}.

Black hole perturbation theory occupies a central position in modern gravitational physics research. The LIGO and Virgo collaboration has successfully detected gravitational wave signals from dozens of black hole merger events through their gravitational wave detector network\cite{LIGOScientific:2016aoc,LIGOScientific:2020zkf}. These observational results not only pioneered the new observational field of gravitational wave astronomy, expanding human means of exploring the universe, but also provided an unprecedented precise verification of GR in strong gravitational field regions. From the characteristic analysis of detected gravitational wave signals, the black hole merger process can be precisely distinguished into three consecutive phases with significantly different physical characteristics: the inspiral phase, merger phase, and ringdown phase\cite{LIGOScientific:2016sjg, LIGOScientific:2017ycc, LIGOScientific:2017vwq}. Among these, the ringdown phase reflects the characteristic oscillation modes of the perturbed black hole, namely the QNMs\cite{Konoplya:2011qq}. These oscillations possess a definite complex frequency characteristic, with its real part representing the oscillation frequency and the imaginary part representing the amplitude decay rate. According to the no-hair theorem\cite{Johannsen:2013szh}, the spectrum of QNMs is completely determined by the fundamental parameters of the black hole (mass, angular momentum, charge, etc.), independent of the specific initial conditions that triggered the perturbation. Therefore, by precisely measuring QNMs, it is theoretically possible to achieve precise constraints on black hole parameters, providing important evidence for testing gravitational theories.Reviewing the historical development of QNMs research, this concept was initially proposed by Vishveshwara during his study of black hole perturbation theory\cite{Vishveshwara:1970zz, Vishveshwara:1970cc}. Subsequently, Schutz and Will were the first to apply the WKB method to calculate black hole QNMs\cite{Schutz:1985km}, and Iyer and Konoplya further improved this method with higher-order corrections, significantly enhancing its accuracy\cite{Konoplya:2011qq, Iyer:1986np}. Leaver proposed a semi-analytical method based on continued fraction expansion, successfully calculating the QNMs spectra for Schwarzschild and Kerr black holes, which is still considered one of the most precise calculation methods\cite{Leaver:1985ax, Leaver:1986gd}. Additionally, Stefanov theoretically proved the mathematical correspondence between black hole QNMs and photon ring structures under strong gravitational lensing conditions, providing a theoretical foundation for studying the connection between black hole perturbations and observational characteristics\cite{Stefanov:2010xz}.Recently, Thomas F.M. Spieksma et al. explored the influence of environments on the ringdown phenomenon that occurs after a binary coalescence\cite{Spieksma:2024voy}.Konoplya et al. investigated the QNMs characteristics of the quantum-corrected Schwarzschild black hole (Bardeen spacetime) using scalar, electromagnetic, and neutrino fields\cite{Konoplya:2023ahd}.Gogoi et al. use the higher-order WKB method in the non-minimal Einstein–Yang–Mills theory and black hole shadows to treat scalar QNMs\cite{Gogoi:2024vcx}.Zinhailo discussed the QNMs of a massive scalar field as the simplest qualitative model for higher spin particles\cite{Zinhailo:2024jzt}.Based on asymptotically safe gravity, Stashko analyzes the QNMs of different fields of regular black hole spacetime consistent with the static exterior of the collapsing dust ball\cite{Stashko:2024wuq}.These results lay the foundation for further research of QNMs.

While delving into black hole QNMs, it is also necessary to consider the influence of the complex material environment surrounding the black hole on its dynamical behavior. Unlike accretion disk matter and conventional celestial bodies that can be directly detected through electromagnetic radiation, dark matter does not participate in electromagnetic interactions and cannot be directly detected by existing astronomical observation methods. Nevertheless, extensive astronomical observational evidence indicates that dark matter plays a decisive role in cosmic large-scale structure formation, galaxy evolution, and cosmic expansion history through gravitational effects\cite{Davis:1985rj, Rubin:1980zd, Turner:1984nf}. Accordingly, studying the distribution patterns, dynamical evolution, and its influence on black hole physical characteristics constitutes an extremely challenging frontier scientific problem in contemporary astrophysics and gravitational theory. Currently, theoretical physicists have proposed multiple dark matter models to explain its cosmological characteristics, primarily including Cold Dark Matter (CDM), Scalar Field Dark Matter (SFDM), Warm Dark Matter (WDM), and PFDM\cite{Blumenthal:1984bp, Hu:2000ke, Bode:2000gq, Potapov:2016obe, Spergel:1999mh}. Among these, the PFDM model has attracted significant attention for its ability to naturally explain galaxy rotation curve flattening and other observational phenomena, providing a theoretical framework for studying dark matter-black hole interactions.In the field of black hole and dark matter interaction research, Liu et al. systematically analyzed the modification effects of dark matter halos on black hole spacetime geometry\cite{Liu:2023vno,Liu:2024xcd,Yang:2023tip,Yang:2022ifo}.Pantig et al. explored the relationship between dark matter and the weak deflection angle of the black hole at the center of the Milky Way\cite{Pantig:2022toh}.Speeney et al. investigated the gravitational wave fluxes produced by different models of dark matter density distribution\cite{Speeney:2024mas}.Gliorio et al. found that LISA can detect the influence of dark matter halos and distinguish between different dark matter halo models\cite{Gliorio:2025cbh}.Xavier et al. considered the effect of dark matter distribution on black hole shadows\cite{Xavier:2023exm}.Ma et al. discuss the properties of Euler-Heisenberg black holes based on PFDM\cite{Ma:2024oqe}.Das et al. studied time-like and null geodesics using charged black holes in the context of PFDM\cite{Das:2023ess}.Navarro et al. conducted high-precision N-body numerical simulations to deeply investigate the structure of dark halos in the standard CDM cosmological model, providing important clues for understanding dark matter's large-scale distribution\cite{Navarro:1995iw}. Klypin et al. based on MultiDark simulations, performed statistical analyses of dark matter halo internal structure, density distribution, and concentration parameters\cite{Klypin:2014kpa}. These research achievements have laid a solid foundation for understanding the complex interaction mechanisms between dark matter and black holes.

In addition, QNMs serve as an important tool for studying black hole stability and dynamical response, with unique advantages: they not only reveal the response characteristics of black holes to external perturbations, but also reflect physical information about the environment surrounding black holes through their spectral features. Based on this advantage, this paper adopts a generalized spacetime metric that simultaneously includes both black hole magnetic charge parameter and PFDM parameter, which was proposed by Amnish Vachher et al\cite{Vachher:2024ldc}. This research utilizes observational data from the EHT for the supermassive black hole M87* at the center of the galaxy to strictly constrain the magnetic charge parameter $a$ and dark matter parameter $\lambda$ in the metric, determining their physically effective parameter space. Subsequently, values are taken within this parameter space to calculate the QNMs spectrum of the black hole under various field perturbations for different parameter combinations, and to analyze the patterns of how parameter changes affect the spectral characteristics.

The structure of this paper is as follows. In Section \ref{2.0}, we introduce PFDM spacetimes and derive the wave equation, calculating its effective potential. In Section \ref{3.0} primarily introduces two methods used in this study: the WKB method and time-domain method. In Section \ref{4.0} ,we calculate the black hole shadow and constrain the parameters$ a$ and $\lambda$ using observational data from the M87* black hole shadow. In Section \ref{5.0}, we study the QNMs of black holes in PFDM and compare them with Schwarzschild black holes. In Section \ref{6.0} presents our conclusions. In this research, Greek indices range from 1 to 4, and we use natural units $G = c = 1$.

\section{PFDM SPACETIMES AND WAVE EQUATION}\label{2.0}
Here we use the spherically symmetric black hole metric surrounded by PFDM derived in the literature\cite{Vachher:2024ldc} by coupling GR with nonlinear electrodynamics(NED)\cite{Ghosh:2021clx,Kumar:2023gjt}, with the derivation process as follows:

First, the coupling action is
\begin{equation}
S=\frac{1}{16 \pi} \int d^{4} x \sqrt{-g}[R-\mathcal{L}(F)+S_{D M},
\end{equation}
where R is the Ricci scalar, $g=\left|g_{bc}\right|$ , $\mathcal{L}(F)$ is the Lagrangian density , related to $F=\frac{1}{4}F_{bc} F^{bc}  $ , $F_{bc}=\partial_{b} A_{c}-\partial_{c}A_{b} $ is the field strength tensor, $ A_{b}$ is the four-potential, $ S_{D M}$ is the dark matter action. By variation of the action
 \begin{equation}
G_{bc} = T_{bc}^{NED} + T_{bc}^{DM} = 2\left[\frac{\partial\mathcal{L}}{\partial F}F_{bd}F_{c}^{d} - g_{bc}\mathcal{L}(F)\right] + T_{bc}^{DM}.
\label{g}
 \end{equation}
 Assuming a static spherically symmetric metric 
\begin{equation}
 d s^{2}=-f(r) d t^{2}+f(r)^{-1} d r^{2}+r^{2}\left(d \theta^{2}+\sin ^{2} \theta d \phi^{2}\right),
\label{1} 
\end{equation}
For the spacetime, the non-zero components of the field strength tensor $ F_{bc}$  are  $F_{01}$  and  $F_{23}$  , and selecting the Maxwell field
\begin{equation}
F_{bc}=2 \delta^{\theta}\left[_\mu \delta_{\nu}^{\phi}\right] a(r) \sin \theta,
\end{equation}
Using the Bianchi identity
 to determine  $dF=0$ , subsequently confirming  $a(r)=c o n s t=a$  , where  $a$  represents the magnetic charge.Obtaining the magnetic field intensity
 \begin{equation}
  F_{23}=2 a \sin \theta, F=\frac{a^{2}}{2 r^{4}},
  \label{c}
 \end{equation}
Selecting the Lagrangian as $ \mathcal{L}(F)=\frac{2 \sqrt{a} F^{5 / 4}}{m(\sqrt{2}+2 g \sqrt{F})^{3 / 2}}$  , where  $m$  is an undetermined constant.
Since spherical symmetry, the PFDM energy-momentum\cite{Zhang:2020mxi} is 
\begin{equation}
 T_{bc}^{D M}=\operatorname{diag}\left(-\mathscr{E}, P_{1}, P_{2}, P_{3}\right), 
 \end{equation} 
 where
 \begin{equation} 
  \mathscr{E}=-P_{1}=-\frac{\lambda}{8 \pi r^{3}} ,
  \label{e}
  \end{equation}
\begin{equation}
P_{2}=P_{3}=-\frac{\lambda}{16 \pi r^{3}}.
\label{p}
\end{equation}
Substituting equations \eqref{1}, \eqref{c}, \eqref{e}, \eqref{p} and \eqref{g}, we obtain the metric function\cite{Vachher:2024ldc}  
\begin{equation}
f(r)=1-\frac{2 M}{\sqrt{r^{2}+a^{2}}}+\frac{\lambda}{r} \log \frac{r}{|\lambda|},
\label{2}
\end{equation}
here $M$ is the black hole mass, $a$ is a parameter related to magnetic charge, $\lambda$ is the PFDM parameter. When $a \to  0$  and $ \lambda  \to  0$, it reduces to the Schwarzschild black hole metric.

\begin{figure}[]
\includegraphics[width=0.5 \textwidth]{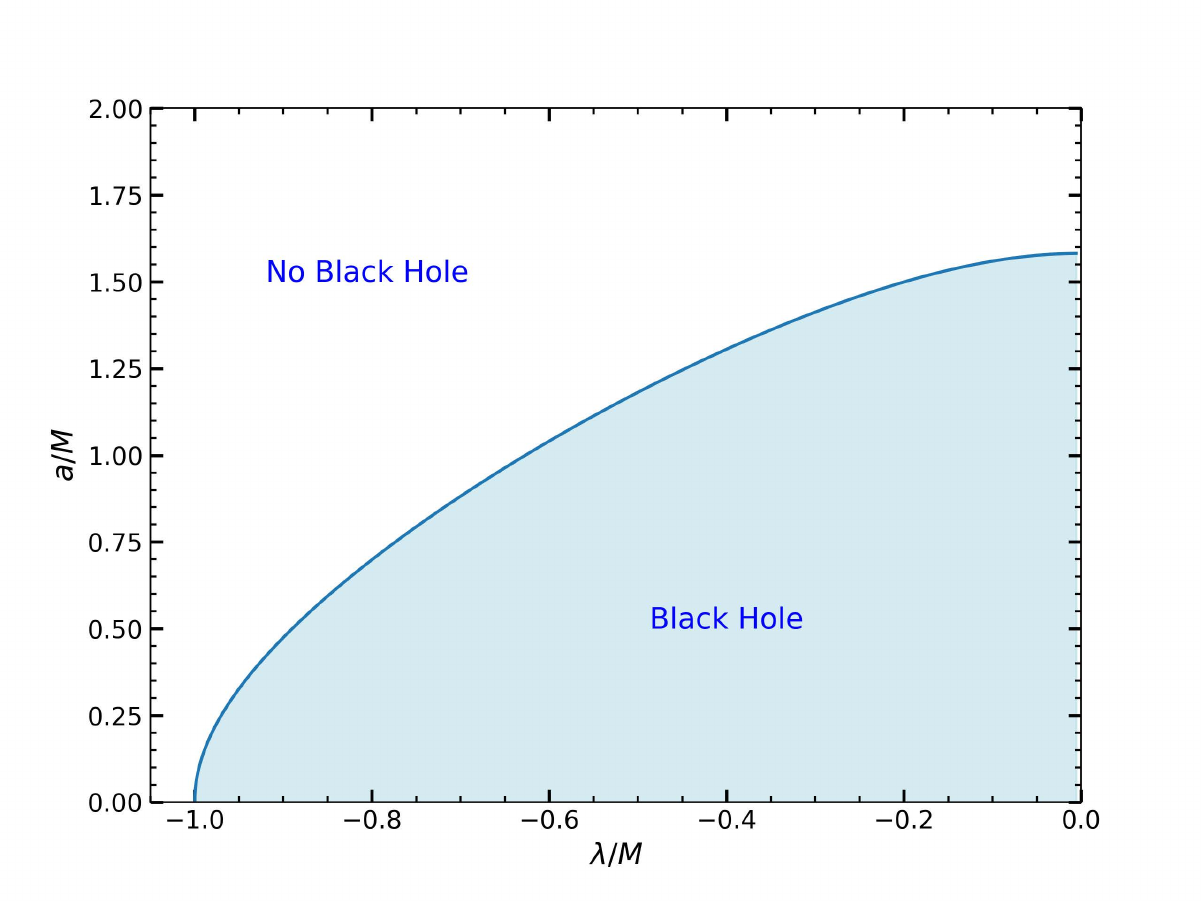}
\captionsetup{justification=raggedright,singlelinecheck=false} 
\caption{
Parameter space ($a/M$, $\lambda/M$) representing a charged black hole surrounded by PFDM. Blue solid line corresponds to extreme black holes.The blue region represents the range of parameter values where black holes exist.}
\label{fig:1}
\end{figure}

\begin{figure*}[htpb]
	\centering
	\includegraphics[width=0.45\textwidth]{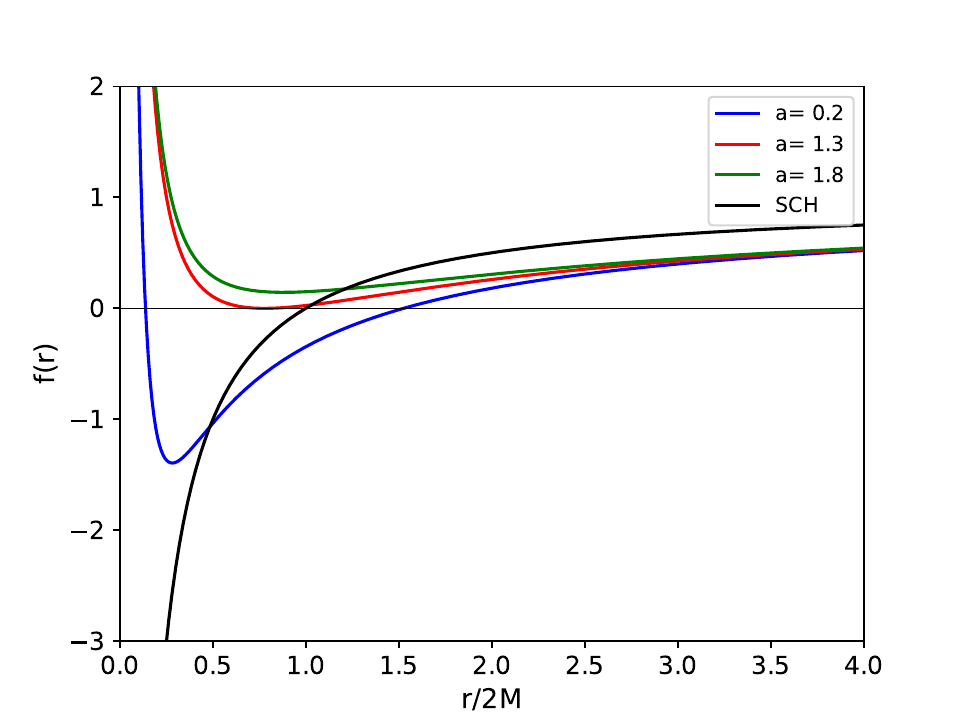}
	\includegraphics[width=0.45\textwidth]{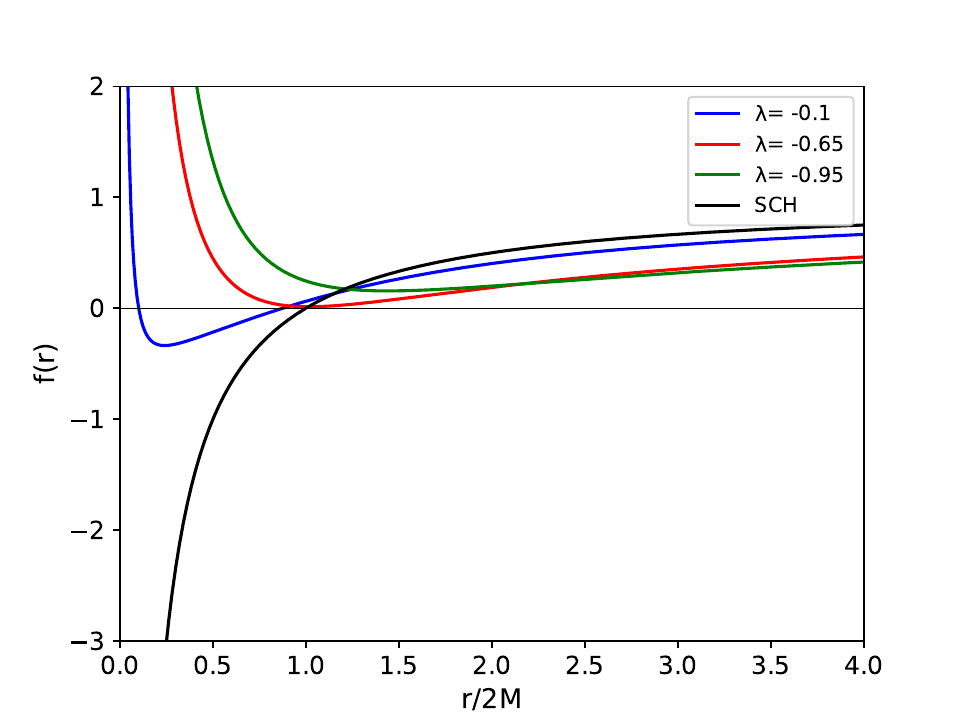}
	\captionsetup{justification=raggedright,singlelinecheck=false} 
	\caption{The horizons of a charged black hole surrounded by PFDM for different $a$ and $\lambda$ parameter values. Black curve is the Schwarzschild black hole horizon, with extreme black hole horizon parameters approximately near the red curve parameters.
	}
	\label{fig:2}	
\end{figure*}

Figure \ref{fig:1} presents the existence region of black hole solutions in the parameter space ($a/M$, $\lambda/M$). The blue solid line characterizes the boundary conditions of extreme black hole solutions, where the inner and outer horizons of the black hole converge, forming a single horizon surface. The light blue region represents the parameter space where conventional black hole solutions exist, with these black holes possessing distinctly separated inner Cauchy horizon and outer event horizon. Figure \ref{fig:2} further illustrates the variation characteristics of the metric function $f(r)$ with radial coordinate $r$, through comparative analysis by selecting typical values in different parameter regions ($a/M$, $\lambda/M$). Specifically, the green curve corresponds to parameter selections in the region without black hole solutions in Figure \ref{fig:1};the red curve approximately characterizes the boundary conditions of extreme black hole solutions; the blue curve represents the metric characteristics of double-horizon black holes. Compared with the classical Schwarzschild black hole solution (which exhibits a strictly monotonically increasing metric function), the black hole solutions in this study demonstrate non-monotonic behavior in certain intervals, first decreasing and then increasing, reflecting the essential impact of parameters ($a/M$, $\lambda/M$) on spacetime structure.Furthermore, for the convenience of subsequent numerical calculations, the values of $a$ and $\lambda$ appearing in this paper are nondimensionalized, i.e.,$a\equiv a/M$,$\lambda\equiv \lambda/M$.

Consider the Klein-Gordon equation for massless scalar and electromagnetic fields in curved spacetime
\begin{equation}
\frac{1}{\sqrt{-g}}\partial_\mu(\sqrt{-g}g^{\mu\nu}\partial_\nu\phi) = 0,
\label{3}
\end{equation}
\begin{equation}
\frac{1}{\sqrt{-g} } \partial \mu (F_{\rho \sigma }g^{\rho \nu }g^{\sigma \mu }   \sqrt{-g}  )=0,
\label{4}
\end{equation}
where $g^{\mu \nu }$, $g^{\rho \nu }$, $g^{\sigma \mu }$ are inverse metric tensors. Substituting equations \eqref{1}, \eqref{2} into \eqref{3}, \eqref{4}, assuming
\begin{equation}
\Psi(t,r,\theta,\phi)=\frac{1}{r}\sum_{l,m}\psi(t,r)Y_{lm}(\theta,\phi),
\end{equation}
we ultimately obtain the wave equation
\begin{equation}
\frac{\partial^{2}}{\partial t^{2}}\psi(t,r)-\frac{\partial^{2}}{\partial r_{*}^{2}}\psi(t,r)+V(r)\psi(t,r)=0,
\end{equation}
Further derivation yields
\begin{equation}
\frac{\mathrm{d}^2 \Psi_{s}  }{\mathrm{d} r_{*}^{2} }+(\omega ^2-V(r)) \Psi_{s}=0.
\label{5}
\end{equation}
Introducing the tortoise coordinate
\begin{equation}
dr_{*} =\frac{dr}{f(r)},
\label{6} 
\end{equation}
The effective potential takes the form
\begin{equation}
V(r)=f(r)\left [ \frac{l(l+1)}{r^2}+(1-s)\frac{f^{'}(r)}{r}\right ], 
\end{equation}
where s is the spin parameter. s=0 represents a scalar field, s=1 represents an electromagnetic field. The effective potential for the scalar field is
\begin{align}
V(r)=&\left ( 1-\frac{2M}{\sqrt{r^{2}+a^{2} } }+ \frac{\lambda }{r} log{\frac{r}{\left |\lambda  \right | } } \right )\nonumber\\
&\left [ \frac{l(l+1)}{r^2}
+\frac{\frac{2Mr}{(r^2+a^2)^{\frac{3}{2} } } -\frac{\lambda }{r^2}log{\frac{r}{\left |\lambda  \right | } }+\frac{\lambda }{r^2} }{r}\right ] ,
\end{align}
The effective potential for the electromagnetic field is
\begin{equation}
V(r)=\left ( 1-\frac{2M}{\sqrt{r^{2}+a^{2} } }+ \frac{\lambda }{r} log{\frac{r}{\left |\lambda  \right | } } \right ) 
 \frac{l(l+1)}{r^2} .
\end{equation}

\begin{figure*}[htpb]
\centering
    \includegraphics[width=0.45\textwidth]{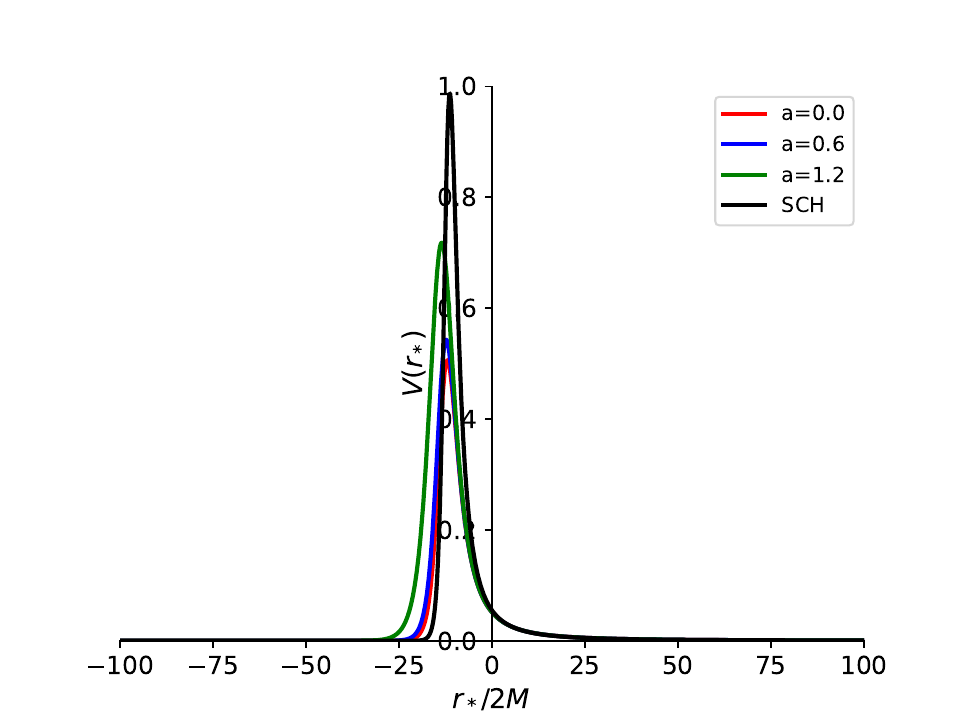}
    \includegraphics[width=0.45\textwidth]{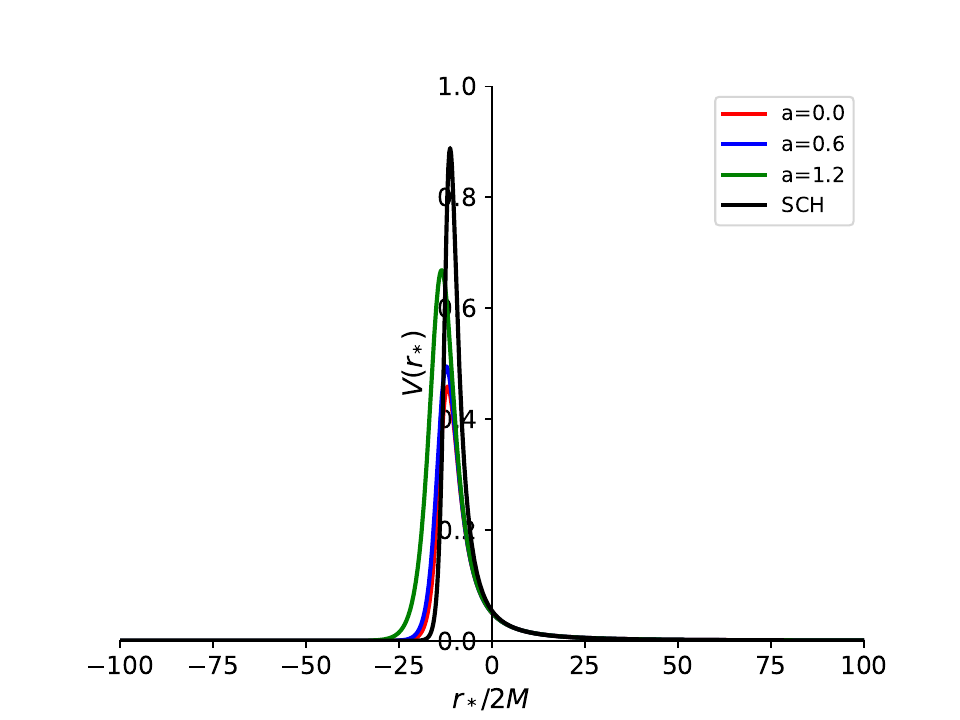}
    \captionsetup{justification=raggedright,singlelinecheck=false}     
\caption{Shows the variation of effective potential energy with tortoise coordinate for scalar field (left) and electromagnetic field (right) in a charged black hole surrounded by ideal fluid dark matter. Set M = 1/2, l = 2, $\lambda$ = -0.15, analyzing effective potential energy changes for different a values.}
\label{fig:3}
\end{figure*}

\begin{figure*}[htpb]
\centering
    \includegraphics[width=0.45\textwidth]{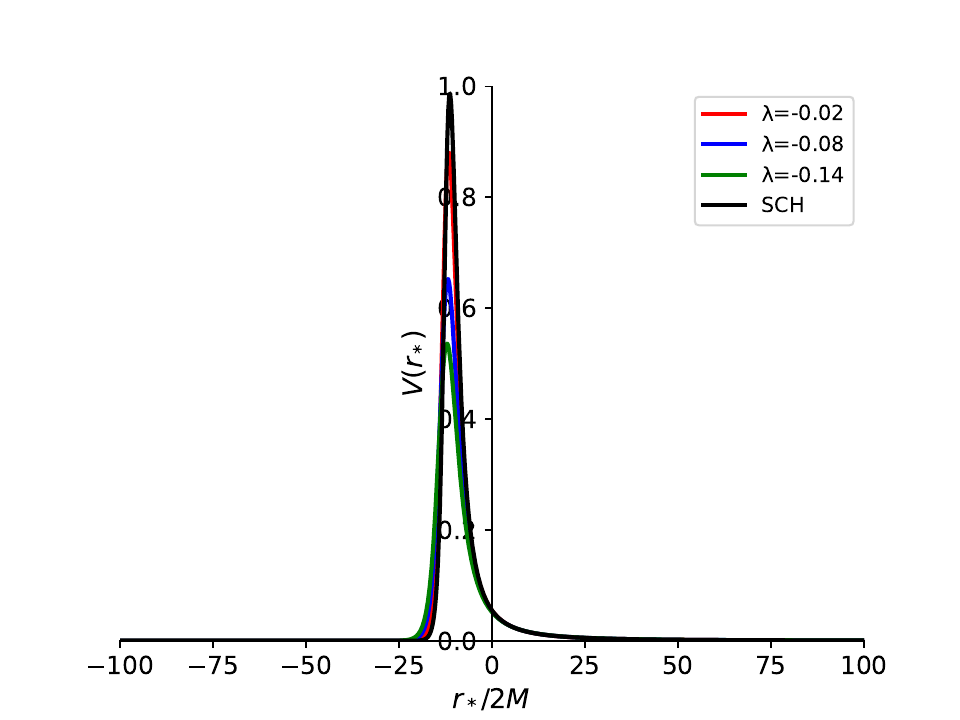}
    \includegraphics[width=0.45\textwidth]{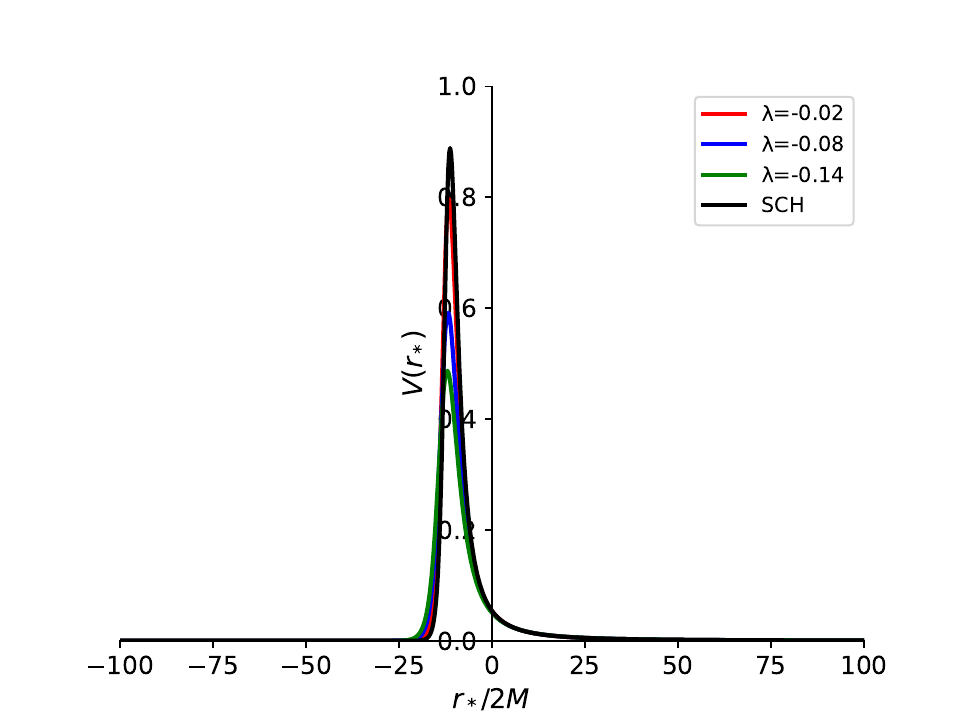}
    \captionsetup{justification=raggedright,singlelinecheck=false} 
\caption{Shows the variation of effective potential energy with tortoise coordinate for scalar field (left) and electromagnetic field (right) in a charged black hole surrounded by ideal fluid dark matter. Set M = 1/2, l = 2, a = 0.4, analyzing effective potential energy changes for different $\lambda$ values.}
\label{fig:4}
\end{figure*}

Figures \ref{fig:3} and Figures \ref{fig:4} present the evolution characteristics of effective potential energy curves under scalar field ($s=0$) and electromagnetic field ($s=1$) conditions in a charged black hole surrounded by ideal fluid dark matter, with comparative analysis against the classical Schwarzschild black hole scenario. These potential energy curves provide crucial insights into understanding black hole stability and dynamical behavior by describing the propagation characteristics of perturbations in the effective potential field.

In Figure \ref{fig:3}, we analyze the impact of magnetic charge parameter $a$ on the potential barrier structure. The results demonstrate that as $a$ increases, the barrier height shows a monotonically increasing trend. This indicates that the increase in magnetic charge parameter $a$ enhances the black hole's resistance to external perturbations. From the perspective of quantum mechanical tunneling effect, higher barriers correspond to lower penetration probabilities, thus making it more difficult for perturbations to penetrate the barrier region. Figure \ref{fig:4} displays the influence of PFDM parameter $\lambda$ on the potential barrier structure. The results reveal that as the absolute value of $\lambda$ increases, the barrier height exhibits a monotonically decreasing trend, which is opposite to the effect of magnetic charge parameter $a$. This phenomenon suggests that an increase in the absolute value of PFDM parameter $\lambda$ reduces the black hole's ``shielding" capacity for perturbations, making it easier for perturbations to penetrate the barrier region. By comparing the left and right panels in both figures, we can clearly observe the significant impact of the perturbation field's spin parameter $s$ on the barrier structure. Under the same parameter configuration, the barrier height for the electromagnetic field ($s=1$) is generally lower than that of the scalar field ($s=0$).

In the context of increased effective potential barrier height, perturbation waves encounter stronger impedance during propagation, leading to a significantly enhanced energy dissipation rate. This accelerated energy dissipation process further induces rapid changes in the spacetime structure near the black hole. Since the spacetime geometric evolution caused by these perturbations directly maps onto the characteristic spectrum of the black hole's QNMs, analyzing the response relationship between perturbations and the black hole's dynamical properties becomes critically important. Therefore, in the next section, we will focus on analyzing two methods for calculating QNMs: the WKB method and the time-domain method.

\section{Method}\label{3.0}
\subsection{WKB Method}\label{3.1}
The WKB (Wentzel-Kramers-Brillouin) method is typically used for solving wave equations, initially introduced by Schutz and Will as a first-order method, and subsequently developed towards higher-order approximations, including 3rd and 6th-order WKB methods\cite{Iyer:1986nq,Konoplya:2003ii}. Matyjasek and Opala extended it to 13th-order using Padé approximation\cite{Matyjasek:2017psv}. It has been applied not only to traditional black hole models but also extended to various complex black hole models including charged, rotating, and higher-dimensional configurations\cite{Konoplya:2003ii,Konoplya:2003dd}. Based on equation \eqref{5}, we can utilize the WKB method to calculate quasinormal frequencies for charged black holes in a PFDM background. The existence of QNMs requires satisfying specific boundary conditions, namely the asymptotic behavior of the wave function at the event horizon and spatial infinity:
\begin{equation}
\Psi\left(r_{*}\right) \sim e^{-i \omega r_{*}}, \quad r_{*}=-\infty,
\end{equation}
\begin{equation}
\Psi\left(r_{*}\right) \sim e^{i \omega r_{*}}, \quad r_{*}=+\infty.
\end{equation}
Here, $r_*$ is the tortoise coordinate, defined by equation \eqref{6}. According to these boundary conditions, only pure ingoing waves exist at the event horizon ($r_* \to -\infty$), while only pure outgoing waves exist at spatial infinity ($r_* \to +\infty$). After applying the WKB method to our model, the quasinormal frequency can be calculated by the following formula:
\begin{equation}
\frac{i(\omega ^{2}-V_{0}  )}{\sqrt{-2{V_{0}}^{''} } } -\sum_{i=2}^{6}\Lambda _{i} =n+\frac{1}{2}  ,  (n=0,1,2,...) ,  
\end{equation}
where $V_{0}$ is the maximum value of the black hole's effective potential, $V_{0}^{\prime\prime}$ is the second-order derivative at the effective potential's maximum point, $\Lambda_{i}$ are the $i$-th order correction terms, and $n$ is the overtone number, with our primary focus on the fundamental mode, i.e., the case of $n=0$.

\subsection{Time-domain Method}\label{3.2}
In the time-domain method, by introducing light-cone coordinates $u =t-r_{*} $, $v =t+r_{*}$, we transform the equation \eqref{5} from conventional coordinates to light-cone coordinates
\begin{equation}
-4\frac{\partial^2 \psi(\mu ,\nu ) }{\partial \mu\partial \nu} =V(\mu ,\nu )\psi(\mu ,\nu ),   
\end{equation}
For numerical implementation, we discretize this partial differential equation, using grid points marked as $N=(u+h,v+h)$, $W=(u+h,v)$, $E=(u,v+h)$, and $S=(u,v)$, where $h$ is the step length of each grid cell. Following the method of Gundlach et al. we adopt a second-order discretization scheme\cite{Gundlach:1993tp,Moderski:2001gt, Moderski:2001tk, Moderski:2005hf}
\begin{align}
\Psi (N)=&\Psi (W)+\Psi (E)-\Psi (S) \nonumber\\
&-h^2\times \frac{V(W)\Psi (W)+V(E)\Psi (E)}{8}+O(h^4) ,  
\end{align}
This scheme is particularly effective because it maintains second-order accuracy while preserving computational efficiency. The error term $O(h^4)$ is generated by the precise handling of mixed derivatives during the discretization process.

A Gaussian pulse is applied on two null surfaces $u=u_{0}$and $v=v_{0}$, expressed as
\begin{equation}
\psi\left(u=u_{0}, v\right)=A \exp \left(-\frac{\left(v-v_{0}\right)^{2}}{2 \sigma^{2}}\right), \quad \psi\left(u, v=v_{0}\right)=0,
\end{equation}
Here $A=1$, $v_{0}=10$, $\sigma=3$. This initial perturbation serves as a trigger for exciting QNMs, enabling us to observe their time evolution.  Finally, we extract QNMs frequencies from the time evolution data using the Prony method\cite{Chowdhury:2020rfj}, with the approximate formula:
\begin{equation}
\psi(t)\simeq \sum_{j=1}^{p}C_{j} e^{-i\omega _{j} t} .    
\end{equation}

\section{Black Hole Shadow Constraints from M87*}\label{4.0}
\begin{figure}[]
\includegraphics[width=0.5 \textwidth]{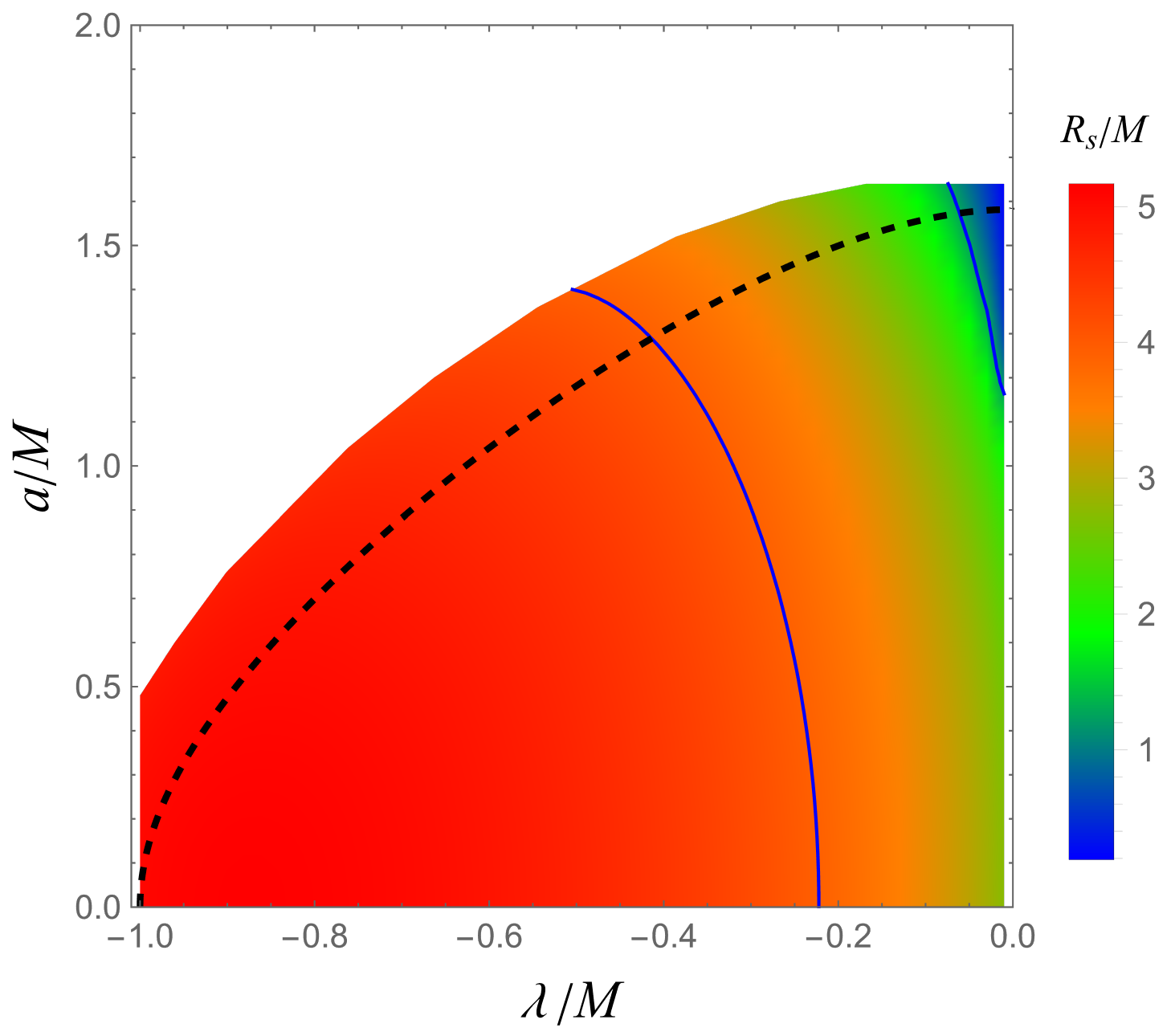} 
\captionsetup{justification=raggedright,singlelinecheck=false} 
\caption{The parameter space constrained by EHT data for charged black holes as M87* candidates in a PFDM background. The black dashed line represents extreme black holes, with the region below the dashed line representing the parameter range where black holes exist, and the space between the two blue curves is constrained by the EHT observations.}
\label{fig:5} 
\end{figure}

In this section, we constrain the parameters of the charged black hole by considering a PFDM background as a massive candidate for M87*, utilizing the EHT observational data of M87*.
First, applying the background metric \eqref{2} yields the Lagrangian equation for geodesics
\begin{widetext}
\begin{equation} 
2\mathcal{L} = g_{\mu\nu}\frac{dx^\mu}{d\chi }\frac{dx^\nu}{d\chi} = -f(r)\left(\frac{dt}{d\chi }\right)^2 + f(r)^{-1}\left(\frac{dr}{d\chi }\right)^2 + r^2\left(\frac{d\theta}{d\chi }\right)^2 + r^2\sin^2\theta\left(\frac{d\varphi}{d\chi }\right)^2, \label{7} \end{equation}
\end{widetext}
where $\chi$ is the affine parameter.
Since the magnetic charge-PFDM black hole is spherically symmetric, without loss of generality, the photon motion can be restricted to the equatorial plane, i.e., $\theta = \pi/2$. Under this condition, equation \eqref{7} can be rewritten as
\begin{equation} 2\mathcal{L} = -f(r)\left(\frac{dt}{d\chi}\right)^2 + f(r)^{-1}\left(\frac{dr}{d\chi}\right)^2 + r^2\left(\frac{d\varphi}{d\chi}\right)^2, \end{equation}
In this process, two conserved quantities emerge:
\begin{equation}
 p_t = \frac{\partial\mathcal{L}}{\partial(\frac{dt}{d\chi})} = -f(r)\frac{dt}{d\chi} = -E, \label{8} 
\end{equation}
\begin{equation}
 p_\varphi=\frac{\partial\mathcal{L}}{\partial(\frac{d\varphi}{d\chi})} = r^2\frac{d\varphi}{d\chi} = L,
\label{9}
\end{equation}
where $E$ and $L$ represent the photon's energy and angular momentum, respectively. Equations \eqref{8} and \eqref{9} can be transformed to
\begin{equation} 
\frac{dt}{d\chi} = \frac{E}{f(r)}, 
\label{10} 
\end{equation}
\begin{equation} 
\frac{d\varphi}{d\chi} = \frac{L}{r^2}. 
\label{11} 
\end{equation}
We primarily consider the null geodesic $\mathcal{L} = 0$, namely
\begin{equation} -f(r)\left(\frac{dt}{d\chi}\right)^2 + f(r)^{-1}\left(\frac{dr}{d\chi}\right)^2 + r^2\left(\frac{d\varphi}{d\chi}\right)^2 = 0, \label{12} \end{equation}
Combining equations \eqref{10}, \eqref{11}, and \eqref{12}, we obtain
\begin{equation} -f(r)\frac{E^2}{f(r)^2} + f(r)^{-1}\left(\frac{dr}{d\chi}\right)^2 + r^2\frac{L^2}{r^4} = 0, \end{equation}
which can be rearranged to
\begin{equation} \left(\frac{dr}{d\chi}\right)^2 = E^2 - \frac{L^2f(r)}{r^2}. \end{equation}
Defining the effective radial potential $U_{\text{eff}}(r) = f(r)/r^2$, circular photon orbits correspond to $dU_{\text{eff}}/dr = 0$, that is
\begin{equation} r_{\text{ps}}f'(r_{\text{ps}}) - 2f(r_{\text{ps}}) = 0, \end{equation}
This equation determines the photon sphere radius $r_{\text{ps}}$. For a distant observer, the black hole shadow's angular radius is
\begin{equation} R_s = \frac{r_{\text{ps}}}{\sqrt{f(r_{\text{ps}})}}. \end{equation}
This relationship establishes a precise correspondence between the photon sphere radius in the theoretical model and the observed black hole shadow angular radius, providing us with a methodological framework for quantitative evaluation of black hole parameters.
We will now use the EHT observed data for M87*: ($2.546M \le R_{s} \le 7.846M$)\cite{Pantig:2024rmr}, to determine the possible allowed parameter range for the charged black hole surrounded by PFDM. Referring to Figure \ref{fig:5}, the space between the two blue curves and the black dashed line is the parameter space constrained by the data observed by the EHT in the case of the black hole under study as a candidate for M87*. From the figure, it can be observed that for lower values of the magnetic charge parameter $a/M$, the effective range of the dark matter parameter $\lambda/M$ is approximately between $-0.2$ and $0$. As the magnetic charge parameter $a/M$ increases, this effective range of $\lambda/M$ gradually shifts to the left, extending toward more negative values,but ultimately can only reach approximately $(-0.5, -0.05)$. In the subsequent research, we will conduct our investigation within this parameter range.

\section{QUASINORMAL MODES}\label{5.0}
This section systematically examines the theoretical characteristics of QNMs complex frequency eigenvalues, where the real part characterizes the intrinsic oscillation frequency of the gravitational system after perturbation, and the imaginary part corresponds to the exponential decay time scale during its relaxation process. Specifically, we conducted an in-depth analysis of the QNMs spectral characteristics of the magnetic charge-PFDM black hole spacetime under the combined action of scalar field perturbations and electromagnetic tensor field perturbations.The study employed WKB method and Prony method to systematically calculate the QNMs frequencies of this type of black hole spacetime under different angular quantum numbers, magnetic charge parameters, and PFDM parameters. To verify the reliability of the computational methods, this research first performed a benchmark test on the QNMs of the classic Schwarzschild spacetime.As shown in Table \ref{tab:1}, the numerical solutions based on sixth-order WKB method and the discrete spectral characteristics extracted by the Prony method are essentially consistent with the Leaver's results \cite{Iyer:1986nq}. Particularly, the different methods demonstrate excellent consistency in low-order mode calculations, which fully confirms the reliability and precision of the numerical scheme employed in this study.

\begin{table*}[]
\centering
\begin{tabular}{>{\centering\arraybackslash}p{2.5cm}>{\centering\arraybackslash}p{4cm}>{\centering\arraybackslash}p{4cm}>{\centering\arraybackslash}p{4cm}}
\hline\hline
\rule{0pt}{12pt}
$l$& Leaver's results&WKB method& Prony method\\
\hline
 \rule{0pt}{11pt}
$l_{SC}=0$ & 0.2210 - 0.2098i & 0.220928 - 0.201638i & 0.235938 - 0.213509 i \\
 \rule{0pt}{11pt}
$l_{SC}=1$ & 0.5858 - 0.1954i & 0.585819 - 0.195523i & 0.584787 - 0.194277 i \\
 \rule{0pt}{11pt}
$l_{SC}=2$ & 0.9672 - 0.1936i & 0.967284 - 0.193532i & 0.966008 - 0.203947 i \\
 \rule{0pt}{11pt}
$l_{EM}=1$ & 0.4966 - 0.1850i & 0.496383 - 0.185274i & 0.497899 - 0.188834 i \\
 \rule{0pt}{11pt}
$l_{EM}=2$ & 0.9152 - 0.1900i & 0.915187 - 0.190022i & 0.914329 - 0.199862 i \\
\hline\hline
\end{tabular}
\captionsetup{justification=raggedright,singlelinecheck=false} 
\caption{Compared the fundamental QNMs of Schwarzschild black holes in scalar and electromagnetic fields using the Prony method, the sixth-order WKB method, and the Leaver's results from the literature \cite{Iyer:1986nq}.}
\label{tab:1}
\end{table*}

\begin{figure*}[htpb]
\centering
    \includegraphics[width=0.45\textwidth]{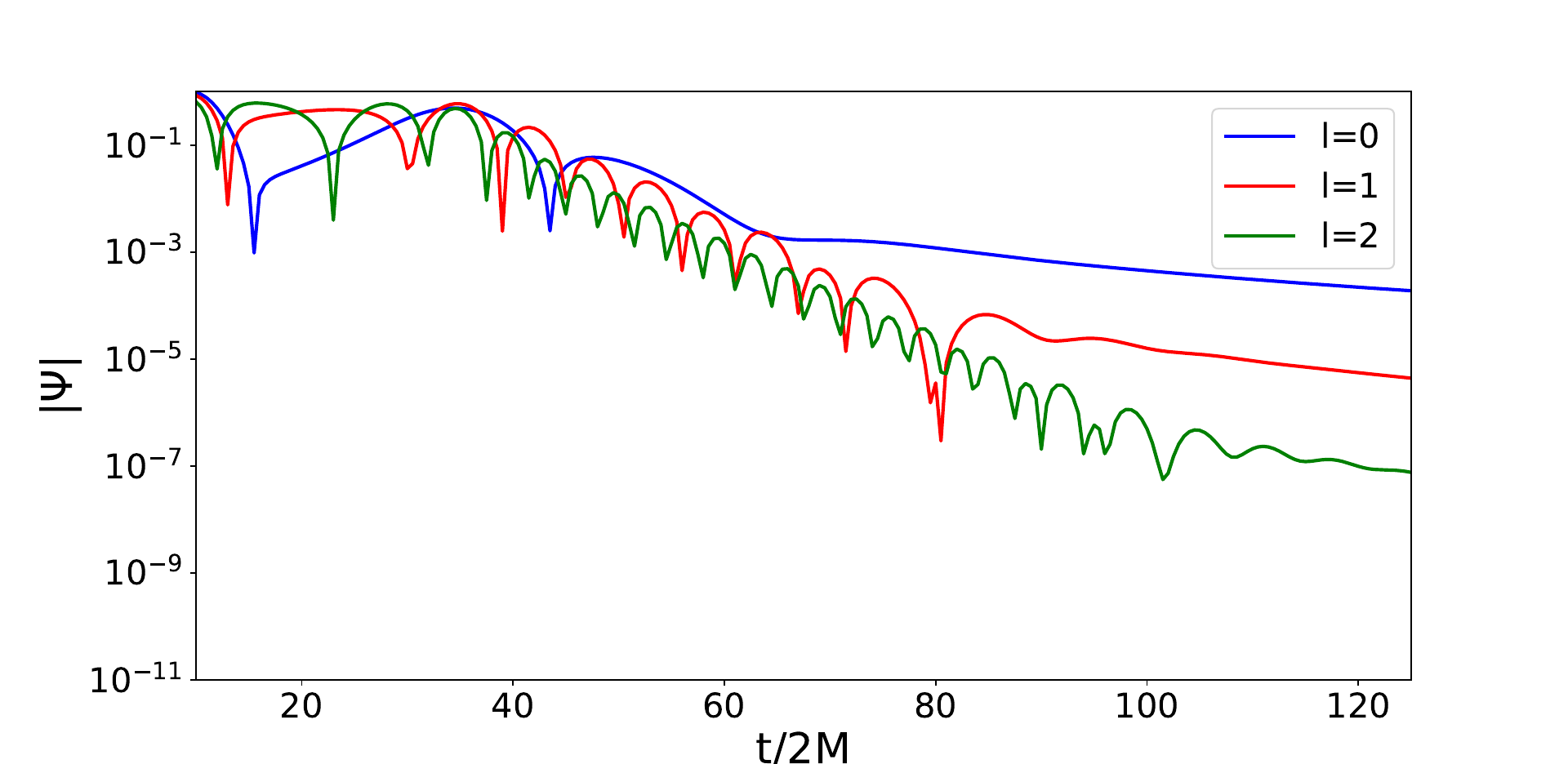}
    \includegraphics[width=0.45\textwidth]{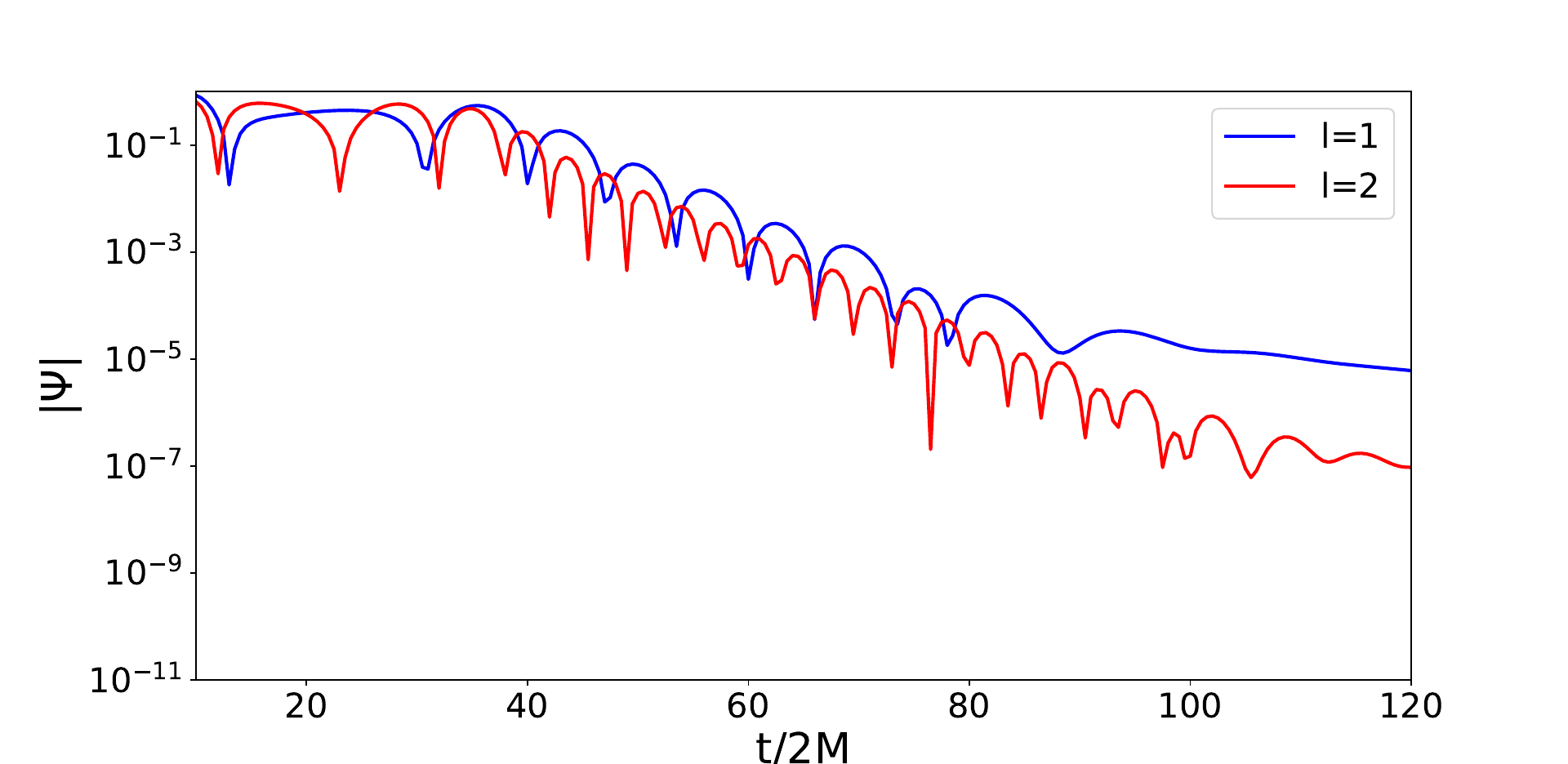} 
\captionsetup{justification=raggedright,singlelinecheck=false} 
\caption{Time evolution of QNMs in scalar field (left) and electromagnetic field (right) perturbations for Schwarzschild black holes, with parameters $M=0.5$.}
\label{fig:6}
\end{figure*}

\begin{figure*}[htpb]
\centering
    \includegraphics[width=0.45\textwidth]{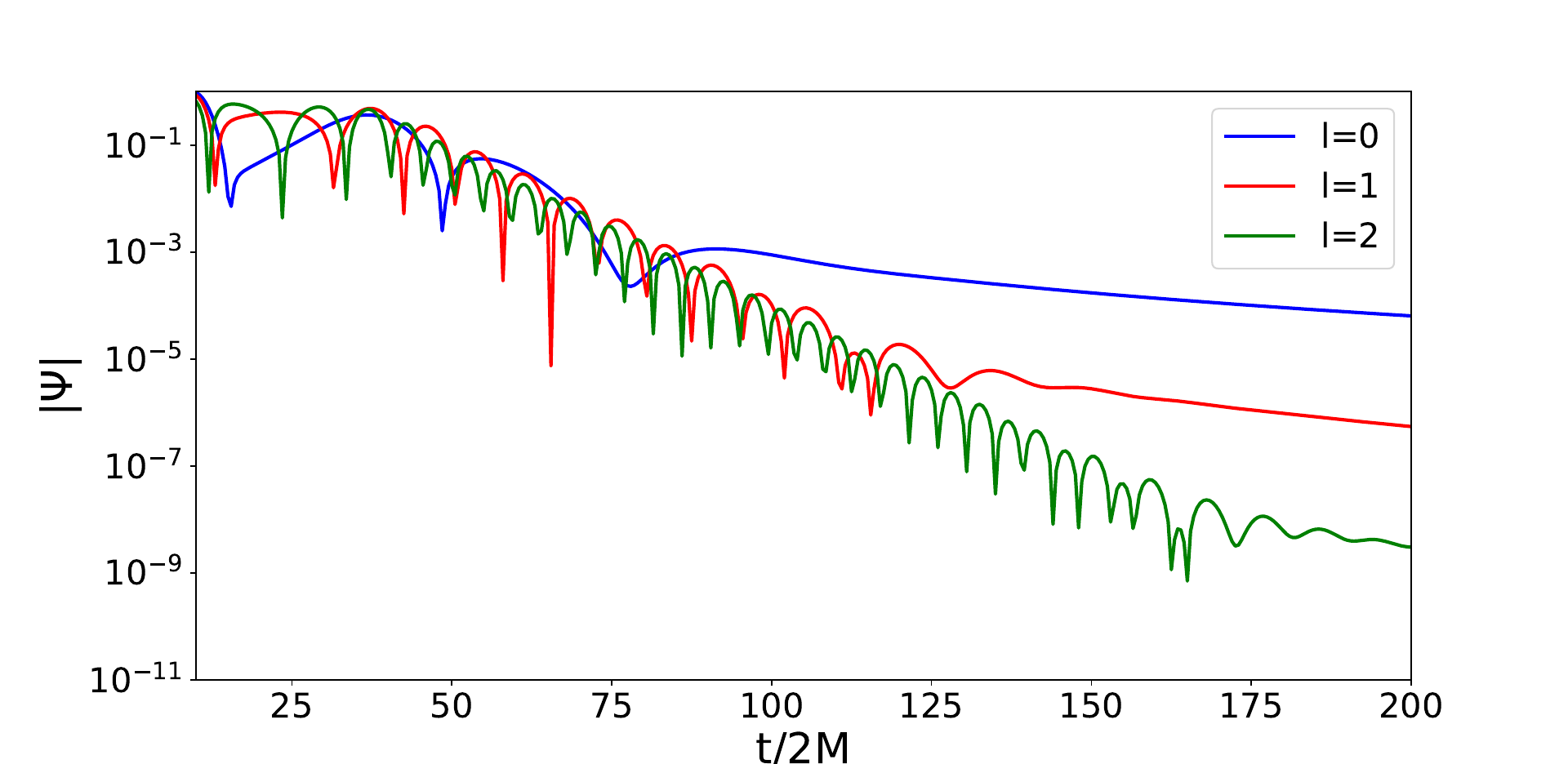}
    \includegraphics[width=0.45\textwidth]{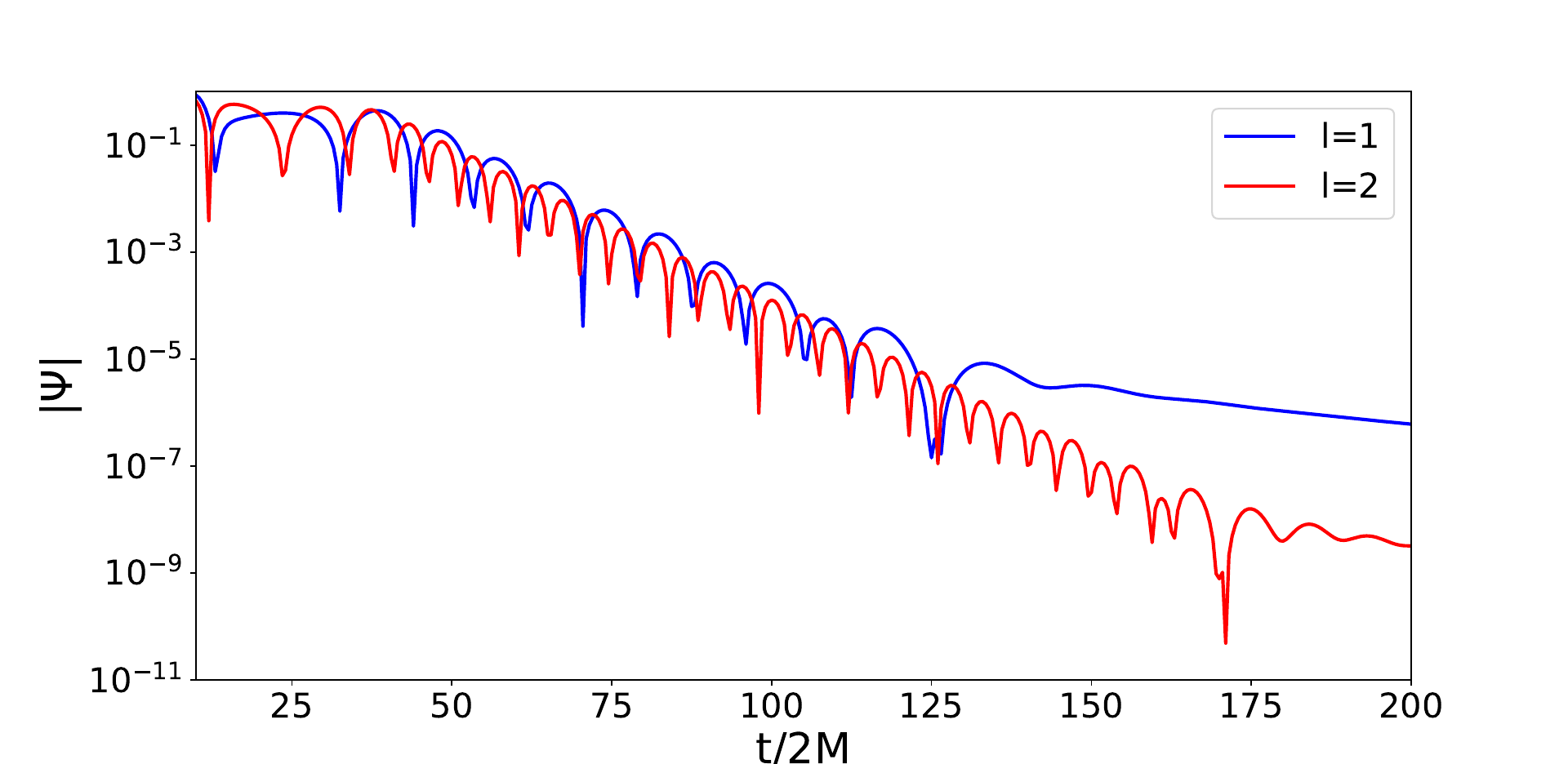} 
\captionsetup{justification=raggedright,singlelinecheck=false} 
\caption{Time evolution of QNMs in scalar field (left) and electromagnetic field (right) perturbations with different angular quantum numbers. Parameters used: $M=0.5$,$ a=0.4$, $\lambda=-0.15$.}
\label{fig:7}
\end{figure*}

\begin{figure*}[htpb]
\centering
    \includegraphics[width=0.45\textwidth]{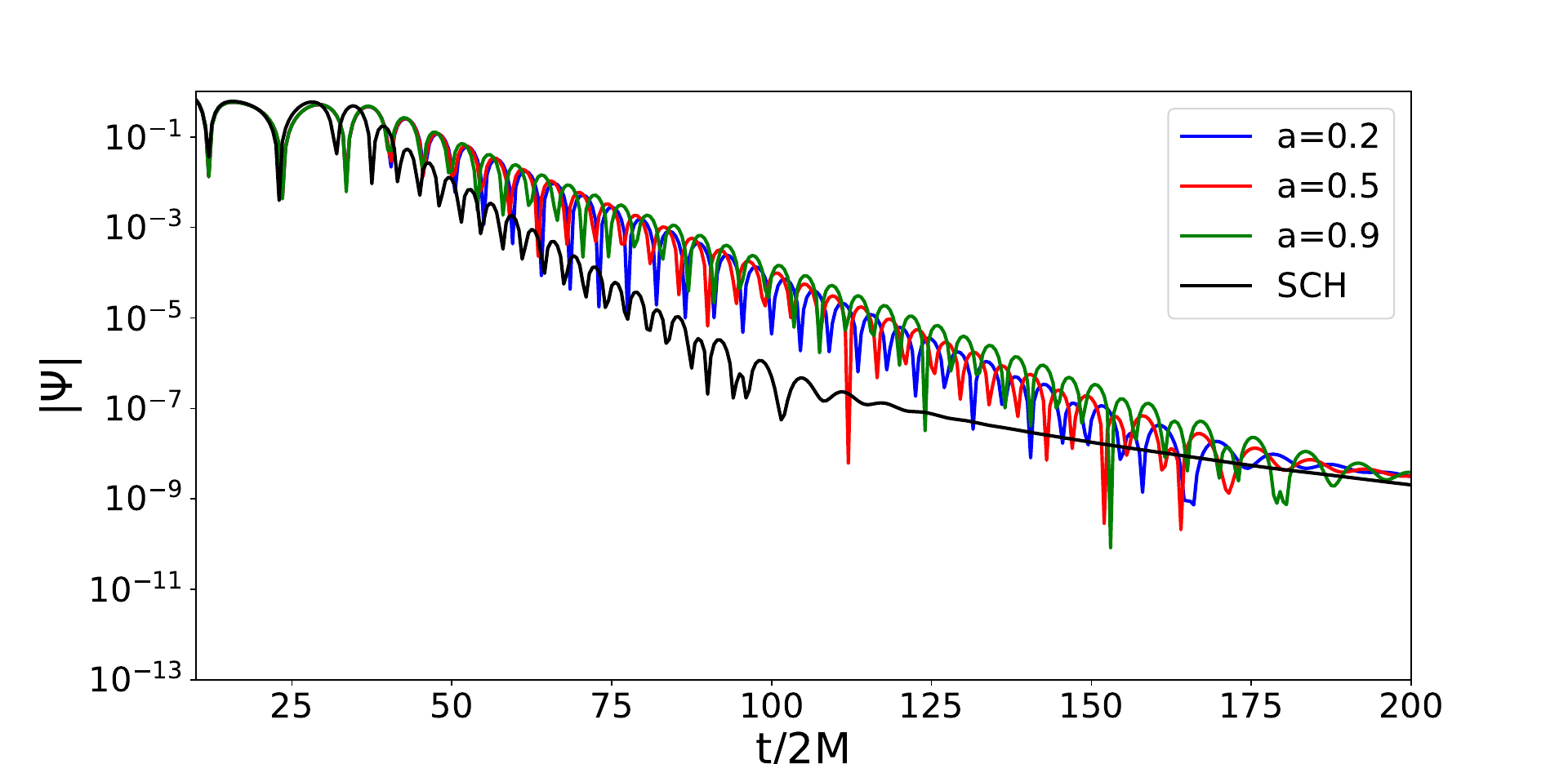}
    \includegraphics[width=0.45\textwidth]{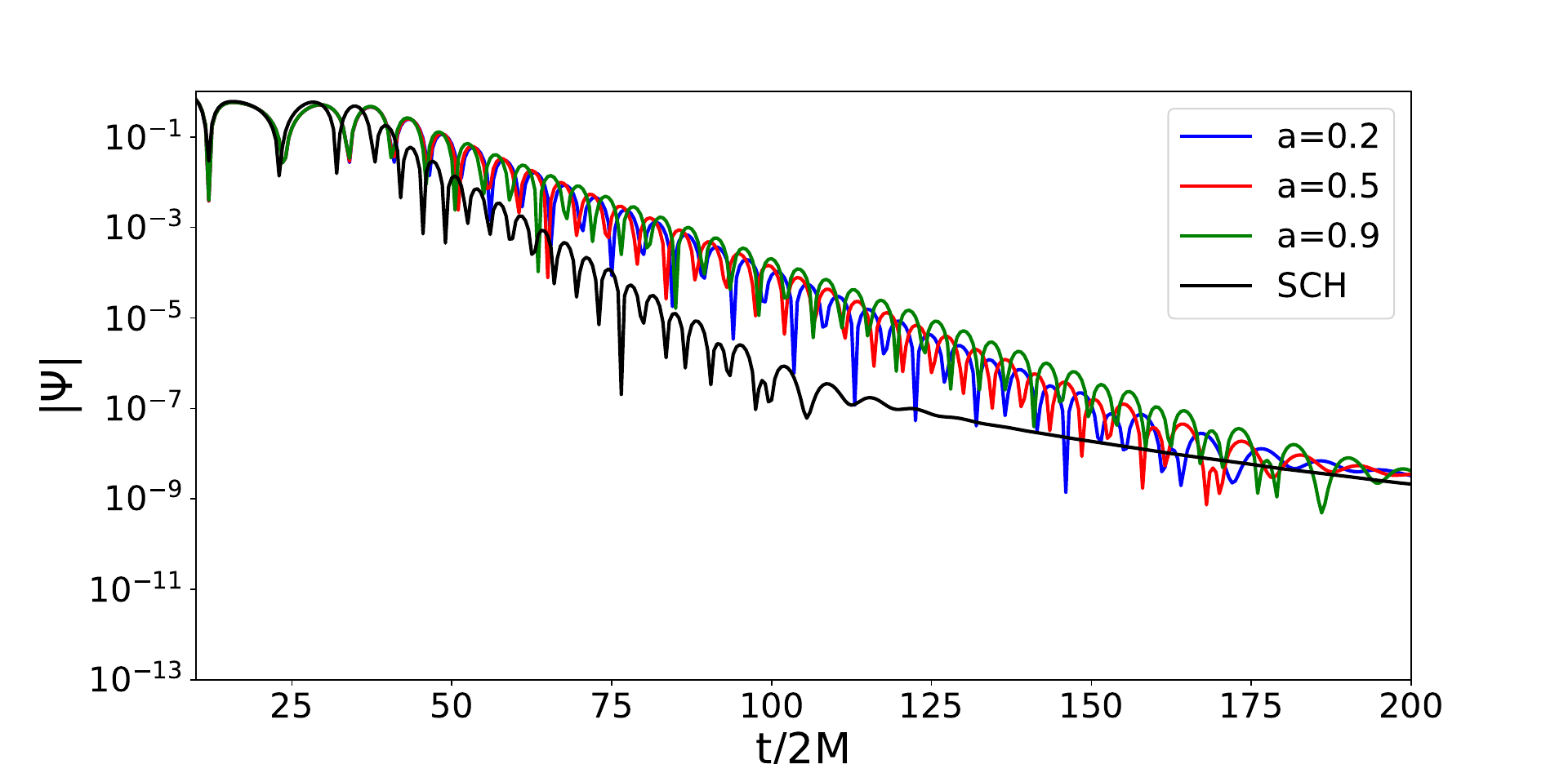}
\captionsetup{justification=raggedright,singlelinecheck=false}  
\caption{Time evolution of QNMs in scalar field (left) and electromagnetic field (right) perturbations with different magnetic charges. Parameters used: $M=0.5$, $\lambda=-0.15$, $l=2$.}
\label{fig:8}
\end{figure*}

\begin{figure*}[htpb]
\centering
    \includegraphics[width=0.45\textwidth]{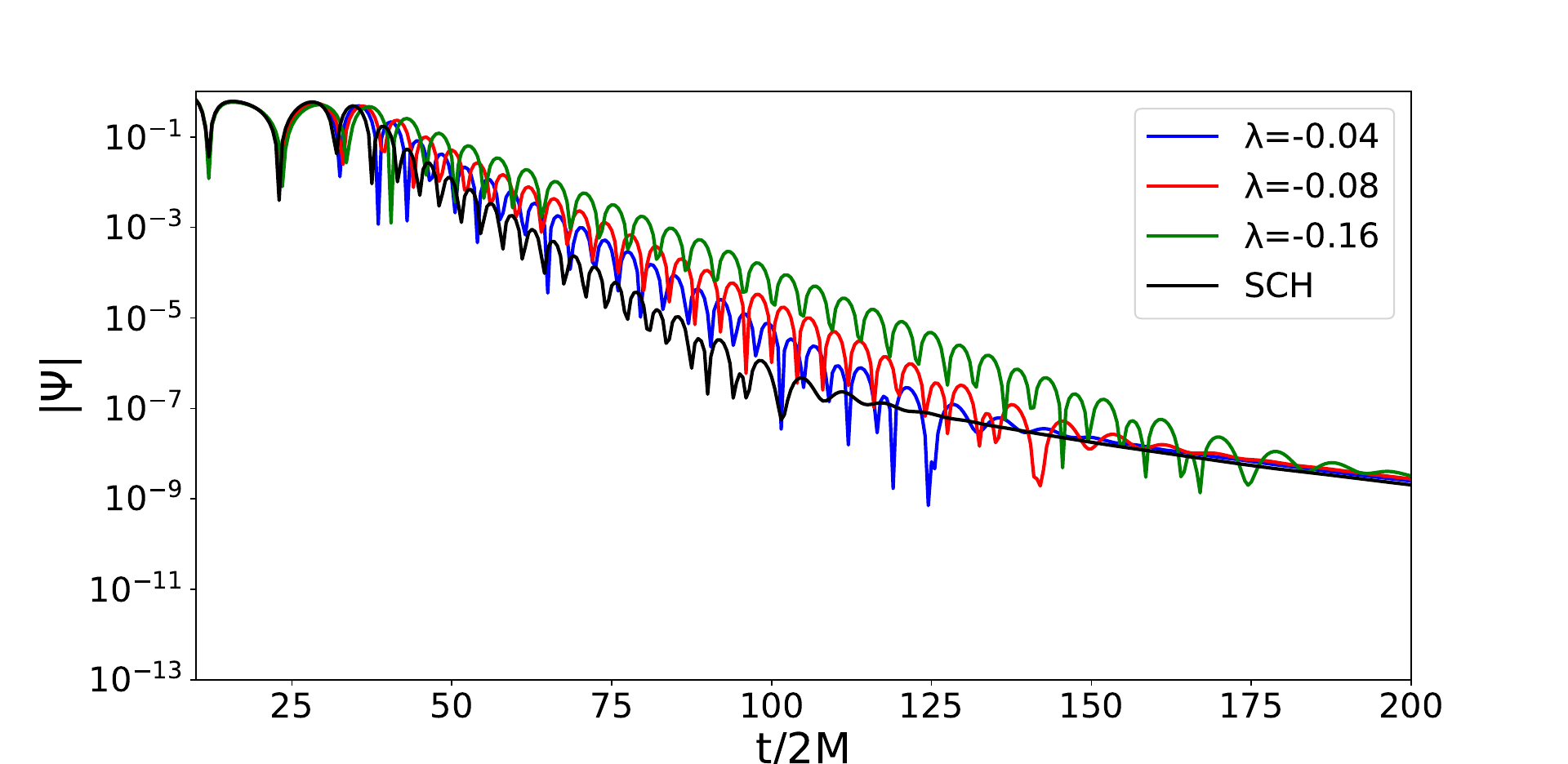}
    \includegraphics[width=0.45\textwidth]{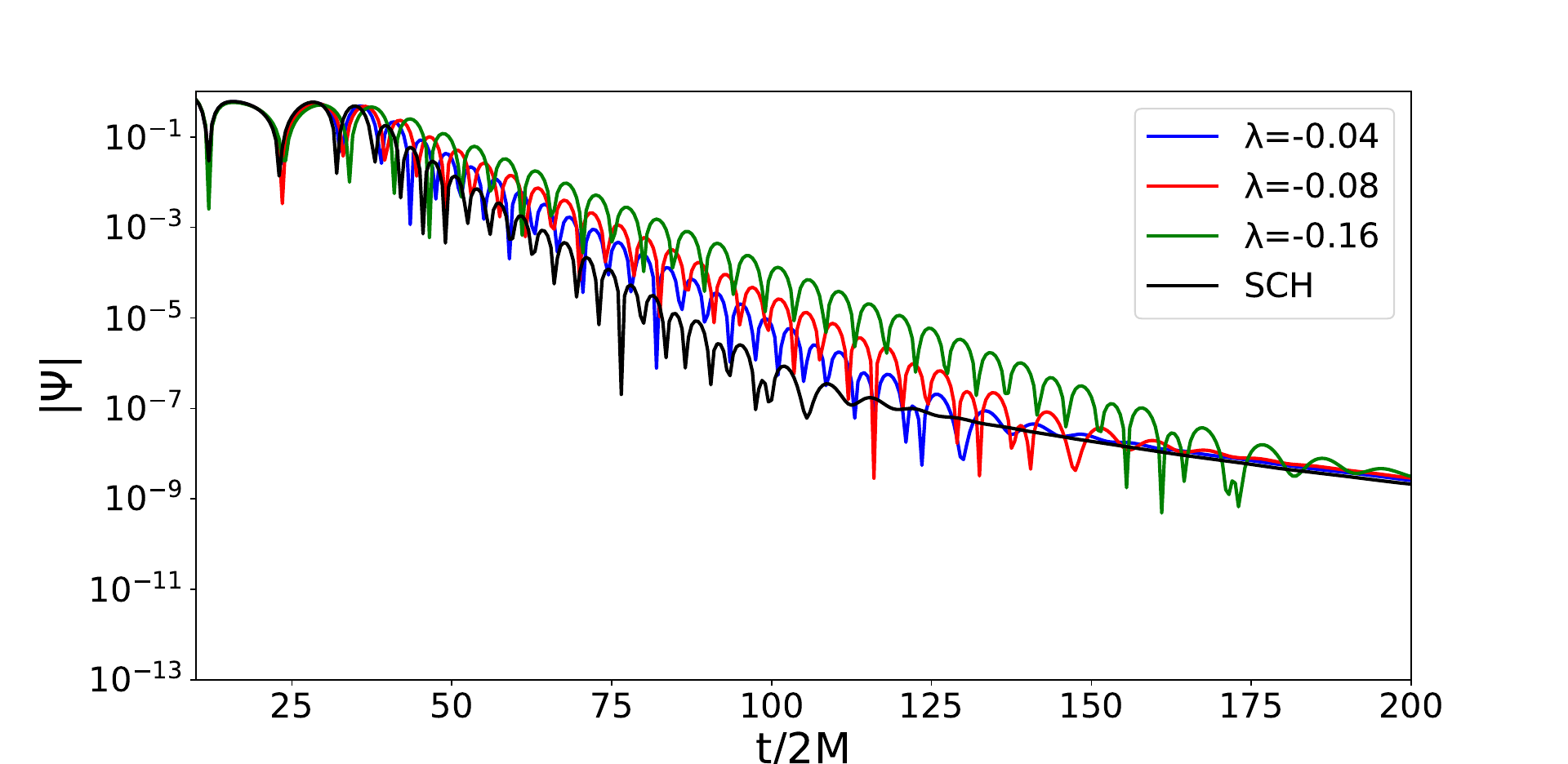}
\captionsetup{justification=raggedright,singlelinecheck=false}  
\caption{Time evolution of QNMs in scalar field (left) and electromagnetic field (right) perturbations with different PFDM parameters. Parameters used: $M=0.5$, $a=0.4$, $l=2$}
\label{fig:9}
\end{figure*}

Figures \ref{fig:7}, Figures\ref{fig:8}, and Figures\ref{fig:9} present the time domain response characteristics of charged PFDM black holes under different perturbation fields. The left and right panels of Figure \ref{fig:7} display the time domain wave profile under scalar field and electromagnetic field perturbations, respectively, with a systematic comparison to the Schwarzschild black hole in Figure \ref{fig:6}. In Figure \ref{fig:7}, we examine the impact of angular quantum number $l$ on the QNMs time domain characteristics.The results demonstrate that as $l$ increases, the signal decay rate significantly strengthens, which corresponds to the increase in the absolute value of the imaginary part of the QNMs frequency. This phenomenon can be physically understood as higher angular quantum number modes possessing higher barrier penetration efficiency, thereby leading to more rapid decay.Figure \ref{fig:8} analyzes the influence of the magnetic charge parameter $a$ on the QNMs. The data shows that when $a$ takes its minimum value, the signal exhibits the strongest decay characteristics. As the $a$ value increases, the signal decay gradually weakens. This implies that the absolute value of the imaginary part of the QNMs frequency decreases with increasing $a$, indicating that the charged effect can slow down the decay process of the perturbation field.Figure \ref{fig:9} investigates the impact of the PFDM parameter $\lambda$ on QNMs characteristics. The results reveal that as $|\lambda|$ increases, the time domain signal decay becomes more gradual, corresponding to a decrease in the absolute value of the QNMs frequency's imaginary part.
\begin{table}[]
\centering
\begin{tabular}{>{\centering\arraybackslash}p{1.5cm}>{\centering\arraybackslash}p{3.5cm}>{\centering\arraybackslash}p{3.5cm}}
\hline\hline
\rule{0pt}{12pt}
$l$&WKB method& Prony method\\
\hline
 \rule{0pt}{11pt}
$l_{SC}=0$&0.160036 - 0.133361i&0.172945 - 0.129040i\\
\rule{0pt}{11pt}
$l_{SC}=1$&0.427310 - 0.130056i&0.427532 - 0.130912i\\
\rule{0pt}{11pt}
$l_{SC}=2$&0.706017 - 0.128996i&0.706516 - 0.131607i\\
\rule{0pt}{11pt}
$l_{EM}=1$&0.368583 - 0.124633i&0.368428 - 0.126633i\\
\rule{0pt}{11pt}
$l_{EM}=2$&0.671733 - 0.127061i&0.671509 - 0.132901i\\
\hline\hline
\end{tabular}
\captionsetup{justification=raggedright,singlelinecheck=false} 
\caption{QNMs frequencies for scalar and electromagnetic fields calculated by WKB method and Prony method under conditions $M=0.5$,$ a=0.4$, $\lambda=-0.15$.}
\label{tab:2}
\end{table}

\begin{table}[]
\centering
\begin{tabular}{>{\centering\arraybackslash}p{1cm}>{\centering\arraybackslash}p{3.5cm}>{\centering\arraybackslash}p{3.5cm}}
\hline\hline
\rule{0pt}{12pt}
$a$&WKB method& Prony method\\
\hline
 \rule{0pt}{11pt}
0.0&0.694375 - 0.130870i&0.694545 - 0.133070i\\
\rule{0pt}{11pt}
0.1&0.695082 - 0.130756i&0.695893 - 0.132588i\\
\rule{0pt}{11pt}
0.2&0.697219 - 0.130412i&0.697535 - 0.132176i\\
\rule{0pt}{11pt}
0.3&0.700835 - 0.129830i&0.700894 - 0.131849i\\
\rule{0pt}{11pt}
0.4&0.706017 - 0.128996i&0.706577 - 0.130323i\\
\rule{0pt}{11pt}
0.5&0.712897 - 0.127889i&0.712038 - 0.129868i\\
\rule{0pt}{11pt}
0.6&0.721659 - 0.126480i&0.721843 - 0.127410i\\
\rule{0pt}{11pt}
0.7&0.732564 - 0.124730i&0.732730 - 0.125141i\\
\rule{0pt}{11pt}
0.8&0.745969 - 0.122586i&0.745689 - 0.122672i\\
\rule{0pt}{11pt}
0.9&0.762369 - 0.119983i&0.762445 - 0.119919i\\
\rule{0pt}{11pt}
1.0&0.782465 - 0.116838i&0.782973 - 0.116525i\\
\rule{0pt}{11pt}
1.1&0.807275 - 0.113051i&0.807490 - 0.113233i\\
\rule{0pt}{11pt}
1.2&0.838323 - 0.108512i&0.837350 - 0.108203i\\
\hline\hline
\end{tabular}
\captionsetup{justification=raggedright,singlelinecheck=false} 
\caption{QNMs frequencies for scalar fields calculated by WKB method and Prony method under conditions$ M=0.5$, $l=2$, $\lambda=-0.15$.}
\label{tab:3}
\end{table}

\begin{table}[]
\centering
\begin{tabular}{>{\centering\arraybackslash}p{1cm}>{\centering\arraybackslash}p{3.5cm}>{\centering\arraybackslash}p{3.5cm}}
\hline\hline
\rule{0pt}{12pt}
$a$&WKB method& Prony method\\
\hline
 \rule{0pt}{11pt}
0.0&0.659757 - 0.128847i&0.659776 - 0.131293i\\
\rule{0pt}{11pt}
0.1&0.660484 - 0.128739i&0.660483 - 0.131174i\\
\rule{0pt}{11pt}
0.2&0.662681 - 0.128411i&0.662657 - 0.130808i\\
\rule{0pt}{11pt}
0.3&0.666401 - 0.127856i&0.666534 - 0.130804i\\
\rule{0pt}{11pt}
0.4&0.671733 - 0.127061i&0.671975 - 0.129035i\\
\rule{0pt}{11pt}
0.5&0.678813 - 0.126004i&0.678926 - 0.127823i\\
\rule{0pt}{11pt}
0.6&0.687835 - 0.124658i&0.687776 - 0.126289i\\
\rule{0pt}{11pt}
0.7&0.699069 - 0.122984i&0.69955 - 0.1241990i\\
\rule{0pt}{11pt}
0.8&0.712887 - 0.120931i&0.712751 - 0.122384i\\
\rule{0pt}{11pt}
0.9&0.729807 - 0.118435i&0.729255 - 0.119511i\\
\rule{0pt}{11pt}
1.0&0.750557 - 0.115420i&0.750848 - 0.114755i\\
\rule{0pt}{11pt}
1.1&0.776193 - 0.111795i&0.776591 - 0.111632i\\
\rule{0pt}{11pt}
1.2&0.808298 - 0.107459i&0.808131 - 0.106520i\\
\hline\hline
\end{tabular}
\captionsetup{justification=raggedright,singlelinecheck=false} 
\caption{QNMs frequencies for electromagnetic fields calculated by WKB method and Prony method under conditions $ M=0.5$, $l=2$, $\lambda=-0.15$.}
\label{tab:4}
\end{table}

\begin{table}[]
\centering
\begin{tabular}{>{\centering\arraybackslash}p{1cm}>{\centering\arraybackslash}p{3.5cm}>{\centering\arraybackslash}p{3.5cm}}
\hline\hline
\rule{0pt}{12pt}
$\lambda$&WKB method& Prony method\\
\hline
 \rule{0pt}{11pt}
-0.02&0.916423 - 0.172340i&0.916246 - 0.180439i\\
\rule{0pt}{11pt}
-0.04&0.863072 - 0.161994i&0.862654 - 0.169313i\\
\rule{0pt}{11pt}
-0.06&0.822362 - 0.153790i&0.820059 - 0.152571i\\
\rule{0pt}{11pt}
-0.08&0.789192 - 0.146923i&0.790169 - 0.146832i\\
\rule{0pt}{11pt}
-0.10&0.761186 - 0.141001i&0.761543 - 0.148941i\\
\rule{0pt}{11pt}
-0.12&0.736988 - 0.135792i&0.736245 - 0.136967i\\
\rule{0pt}{11pt}
-0.14&0.715733 - 0.131144i&0.715939 - 0.132693i\\
\rule{0pt}{11pt}
-0.16&0.696831 - 0.126950i&0.696540 - 0.129825i\\
\rule{0pt}{11pt}
-0.18&0.679861 - 0.123132i&0.679917 - 0.127976i\\
\rule{0pt}{11pt}
-0.20&0.664506 - 0.119631i&0.664387 - 0.123844i\\
\hline\hline
\end{tabular}
\captionsetup{justification=raggedright,singlelinecheck=false} 
\caption{QNMs frequencies for scalar fields calculated by WKB method and Prony method under conditions $ M=0.5$, $l=2$, $a=0.4$.}
\label{tab:5}
\end{table}

\begin{table}[]
\centering
\begin{tabular}{>{\centering\arraybackslash}p{1cm}>{\centering\arraybackslash}p{3.5cm}>{\centering\arraybackslash}p{3.5cm}}
\hline\hline
\rule{0pt}{12pt}
$\lambda$&WKB method& Prony method\\
\hline
 \rule{0pt}{11pt}
-0.02&0.869807 - 0.169467i&0.869492 - 0.172053i\\
\rule{0pt}{11pt}
-0.04&0.819412 - 0.159333i&0.818528 - 0.166271i\\
\rule{0pt}{11pt}
-0.06&0.781040 - 0.151304i&0.780282 - 0.157566i\\
\rule{0pt}{11pt}
-0.08&0.749825 - 0.144587i&0.749392 - 0.145967i\\
\rule{0pt}{11pt}
-0.10&0.723501 - 0.138796i&0.723965 - 0.139438i\\
\rule{0pt}{11pt}
-0.12&0.700779 - 0.133703i&0.700746 - 0.135195i\\
\rule{0pt}{11pt}
-0.14&0.680840 - 0.129160i&0.680555 - 0.132100i\\
\rule{0pt}{11pt}
-0.16&0.663126 - 0.125062i&0.663593 - 0.127963i\\
\rule{0pt}{11pt}
-0.18&0.647233 - 0.121331i&0.647241 - 0.124275i\\
\rule{0pt}{11pt}
-0.20&0.632866 - 0.117911i&0.632145 - 0.119310i\\
\hline\hline
\end{tabular}
\captionsetup{justification=raggedright,singlelinecheck=false} 
\caption{QNMs frequencies for electromagnetic fields calculated by WKB method and Prony method under conditions $ M=0.5$, $l=2$, $a=0.4$.}
\label{tab:6}
\end{table}

In addition to time domain response analysis, we directly calculated the QNMs complex frequencies using frequency domain methods. Table \ref{tab:2} demonstrates the influence of different angular quantum numbers on QNMs frequencies under scalar and electromagnetic field perturbations. Data analysis clearly indicates that as the angular quantum number increases, the real part of the QNMs frequency monotonically increases, and the absolute value of the imaginary part correspondingly increases. This suggests that higher angular quantum number modes exhibit faster oscillation frequencies and stronger decay characteristics, consistent with the predictions of classical black hole perturbation theory.Tables \ref{tab:3} through \ref{tab:6} systematically present the impacts of magnetic charge and PFDM parameters on QNMs frequencies. The data in Tables \ref{tab:3} and \ref{tab:4} reveal that as the magnetic charge parameter $a$ increases, the real part of the QNMs frequency shows a strictly monotonic increasing trend, while the absolute value of the imaginary part strictly monotonically decreases. This implies that stronger electromagnetic effects lead to higher oscillation frequencies but slower signal decay. Physically, this can be explained by the electromagnetic field's modification of the spacetime geometric structure near the black hole horizon, altering the effective potential barrier shape and consequently influencing the QNMs characteristics.Tables \ref{tab:5} and \ref{tab:6} show that as the PFDM parameter $|\lambda|$ increases, both the real part of the QNMs frequency and the absolute value of its imaginary part exhibit a pronounced monotonic decrease. This indicates that dark matter effects reduce the oscillation frequency of the perturbation field and slow its decay process. From a physical mechanism perspective, this may originate from the systematic modification of the black hole spacetime structure by dark matter, particularly the reduction of the effective potential barrier height, which diminishes the radiation efficiency of the perturbation field.
Notably, the two different perturbation fields exhibit similar qualitative behavior in parameter dependence, though with quantitative differences. These frequency domain analysis results are highly consistent with the time domain analysis, mutually validating the computational results and theoretical predictions of this study, and providing a comprehensive and systematic description of the QNMs characteristics of charged PFDM black holes.

\section{Summary}\label{6.0}
In this study, we conducted a systematic investigation of QNMs for charged black holes embedded in a PFDM environment. By utilizing the Event Horizon Telescope (EHT) observational data of the supermassive M87* black hole shadow, we rigorously established the physical constraint conditions for the magnetic charge parameter $a$ and the PFDM density parameter $\lambda$, ensuring our exploration remained within the physically feasible parameter space. Our analysis employed complementary methodological approaches: a  sixth-order WKB  method for frequency calculations, and a time-domain method for perturbation field evolution simulation, while simultaneously using the Prony  method to extract QNMs frequencies from time-domain profiles. The basic consistency between these independent methods validated the reliability and robustness of our computational framework.

The results show that when we select lower magnetic charge parameter values, the PFDM parameter can range approximately from $(-0.2, 0)$; as the chosen magnetic charge parameter increases, the range of acceptable PFDM parameter values shifts toward more negative values, but ultimately can only reach approximately $(-0.5, -0.05)$. Furthermore,for both scalar field and electromagnetic field perturbations, the real part of QNMs frequencies increases with the magnetic charge parameter $a$ and multipolar angular momentum $l$, while decreasing with the increase of PFDM parameter $|\lambda|$. This phenomenon indicates that parameters $a$ and $l$ can effectively enhance the system's oscillatory behavior, whereas parameter $|\lambda|$ produces the opposite effect. Regarding the frequency imaginary part, its absolute value increases with the increase of $l$, suggesting that higher angular momentum accelerates the perturbation signal decay process. Conversely, as $a$ and $|\lambda|$ increase, the absolute value of the imaginary part decreases, implying that stronger dark matter environments and black hole electromagnetic characteristics would decelerate the perturbation field decay process.

Our research results quantitatively describe the coupling effects of black hole magnetic charge parameters and PFDM density parameters in perturbation field propagation.  Particularly, we demonstrated the sensitivity of black hole internal structure to environmental parameter variations, offering a viable diagnostic tool for future gravitational wave detectors to distinguish between different black hole models.

\section*{Acknowledgments}
This research was partly supported by the National Natural Science Foundation of China (Grant No. 12265007).


\bibliography{ref}
\bibliographystyle{apsrev4-1}

\end{document}